\documentclass[12pt]{iopart}
\usepackage{graphicx, color,iopams}
\usepackage[pdftex, bookmarks, colorlinks=true, bookmarksnumbered=true, pdfstartview=FitH, pdfpagemode=UseOutlines, pdfstartpage=1,bookmarksopen=true]{hyperref}
\hypersetup{colorlinks,citecolor=blue,filecolor=blue,linkcolor=blue,urlcolor=blue, pdftex}

\begin{document}

\title[The Path to Enhanced and Advanced LIGO]{The Path to the Enhanced and Advanced LIGO Gravitational-Wave Detectors}


\author{J R Smith$^1$ for the LIGO Scientific Collaboration}

\address{$^1$Department of Physics, Syracuse University, Syracuse, NY USA 13244-1130}

\begin{abstract}
We report on the status of the Laser Interferometric Gravitational-Wave Observatory (LIGO) and the plans and progress towards Enhanced and Advanced LIGO. The initial LIGO detectors have finished a two year long data run during which a full year of triple-coincidence data was collected at design sensitivity. Much of this run was also coincident with the data runs of interferometers in Europe, GEO600 and Virgo. The joint analysis of data from this international network of detectors is ongoing. No gravitational wave signals have been detected in analyses completed to date. Currently two of the LIGO detectors are being upgraded to increase their sensitivity in a program called Enhanced LIGO. The Enhanced LIGO detectors will start another roughly one year long data run with increased sensitivity in 2009. In parallel, construction of Advanced LIGO, a major upgrade to LIGO, has begun. Installation and commissioning of Advanced LIGO hardware at the LIGO sites will commence at the end of the Enhanced LIGO data run in 2011. When fully commissioned, the Advanced LIGO detectors will be ten times as sensitive as the initial LIGO detectors. Advanced LIGO is expected to make several gravitational wave detections per year. 
\end{abstract}
\submitto{\CQG}
\maketitle

\begin{centering}
\begin{figure}[ht]
\includegraphics[width=0.49\textwidth]{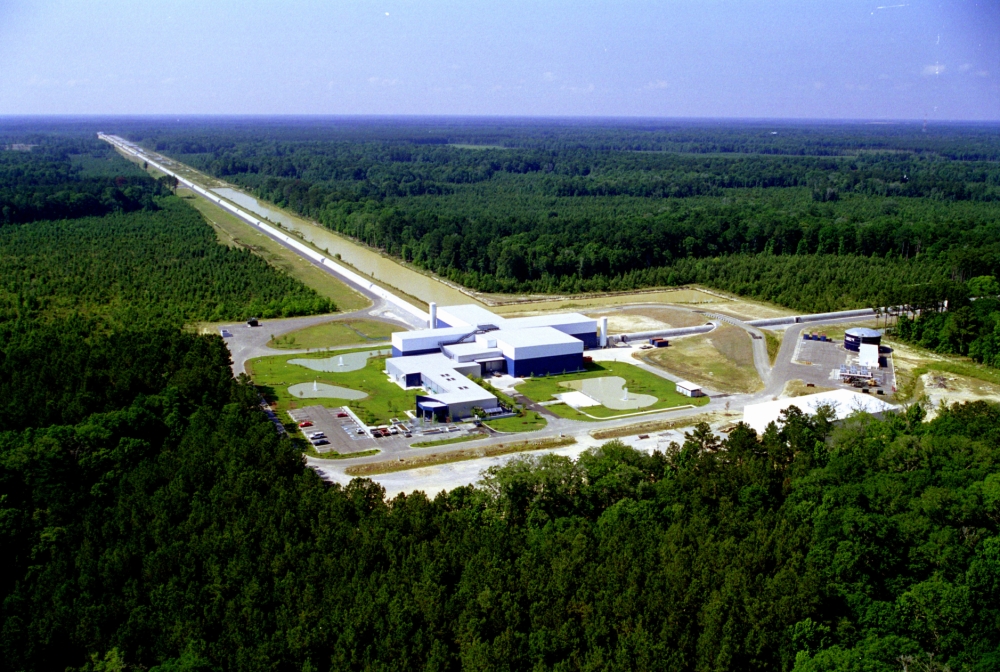}
\includegraphics[width=0.505\textwidth]{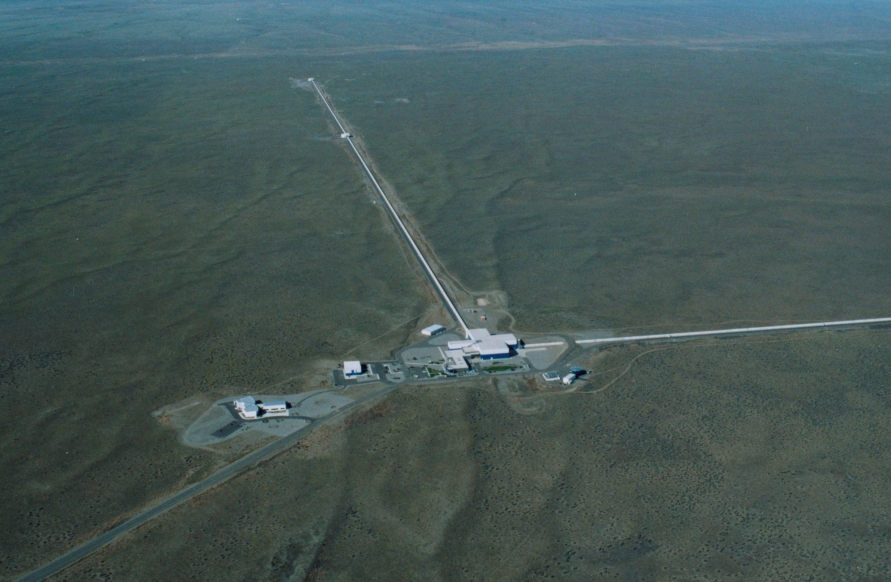}
\caption{The LIGO sites at Livingston, LA (left) and Hanford, WA (right). Photo credit: the LIGO Laboratory.}
\label{fig:ligos}
\end{figure}
\end{centering}

The Laser Interferometric Gravitational-Wave Observatory (LIGO) is a project to detect and study gravitational waves from astrophysical sources that radiate in the audio-frequency regime~\cite{ligo_abrom}. LIGO comprises three interferometers at two sites. The Hanford, WA site has two co-located and co-aligned interferometers, one with 4\,km long arms and one with 2\,km long arms, referred to as H1 and H2, respectively. The Livingston, LA site has one interferometer with 4\,km arms called L1. The LIGO observatories are shown in Figure~\ref{fig:ligos}.

In late 2007 the LIGO detectors completed a two-year long data taking run, the fifth ``science run'' for LIGO, S5, during which one year of triple-coincident science-quality data was collected at design sensitivity. The configuration and sensitivity of the initial LIGO detectors is described in~\cite{ligo, ligo2} and Section~\ref{secn:iligo}. During S5 LIGO was joined in observation by the French-Italian Virgo~\cite{virgo} and UK-German GEO600~\cite{geo} detectors, forming the most sensitive worldwide network of gravitational-wave detectors to date. 

A number of searches for gravitational waves from various sources have already been conducted with data from LIGO and its international partners, see e.g., \cite{cbc, burst, cw, stochastic, burstligogeo, virgoc7}. Data from LIGO, GEO600 and Virgo from S5 (and Virgo's corresponding run, VSR1) is currently being analyzed jointly by the LIGO Scientific Collaboration (which includes the GEO Collaboration) and the Virgo Collaboration. Notable astrophysical results from searches already conducted include a statement that GRB070201 was unlikely to have originated from the merger of a binary system of neutron stars in M31~\cite{s5grb}, and an upper limit on the gravitational radiation emitted by the Crab Pulsar that is significantly lower than the upper limit due to the pulsar's spin-down~\cite{s5crab}. 

After the completion of S5, the two LIGO detectors with 4\,km arms, H1 and L1, were taken offline to receive a number of incremental upgrades that, based on knowledge of the noise sources limiting initial LIGO, will improve their sensitivity by roughly a factor of two. This program is referred to as Enhanced LIGO and is described in~\cite{eligo} and Section~\ref{secn:eligo}. The third detector, H2, was left on-line, in tandem with GEO600 to observe until the upgrades are complete, in a program called Astrowatch. The installation and commissioning required for Enhanced LIGO are nearly complete as of February 2009. LIGO is planning to bring L1 and H1 back online, which together with an upgraded Virgo, will undertake another long data-taking run beginning in mid-2009. 

In parallel with these activities, construction of Advanced LIGO has begun. Advanced LIGO will use the initial LIGO buildings and vacuum systems but will otherwise consist of completely new instruments. All three interferometers will be upgraded and the arms of H2 will be extended to 4\,km. A description of the configuration and expected sensitivity of Advanced LIGO is given in~\cite{aligo} and Section~\ref{secn:aligo}. With a factor of 10 expected sensitivity improvement with respect to initial LIGO, Advanced LIGO will be sensitive enough to record several gravitational wave signals per year from e.g., coalescing binary systems of neutron stars.  Although initial and Enhanced LIGO are sensitive enough to possibly make the first gravitational wave detections, gravitational wave astronomy in the audio-frequency regime will begin in earnest with Advanced LIGO.

\section{Initial LIGO}\label{secn:iligo}


\begin{figure}\begin{center}
\includegraphics[width=1\textwidth]{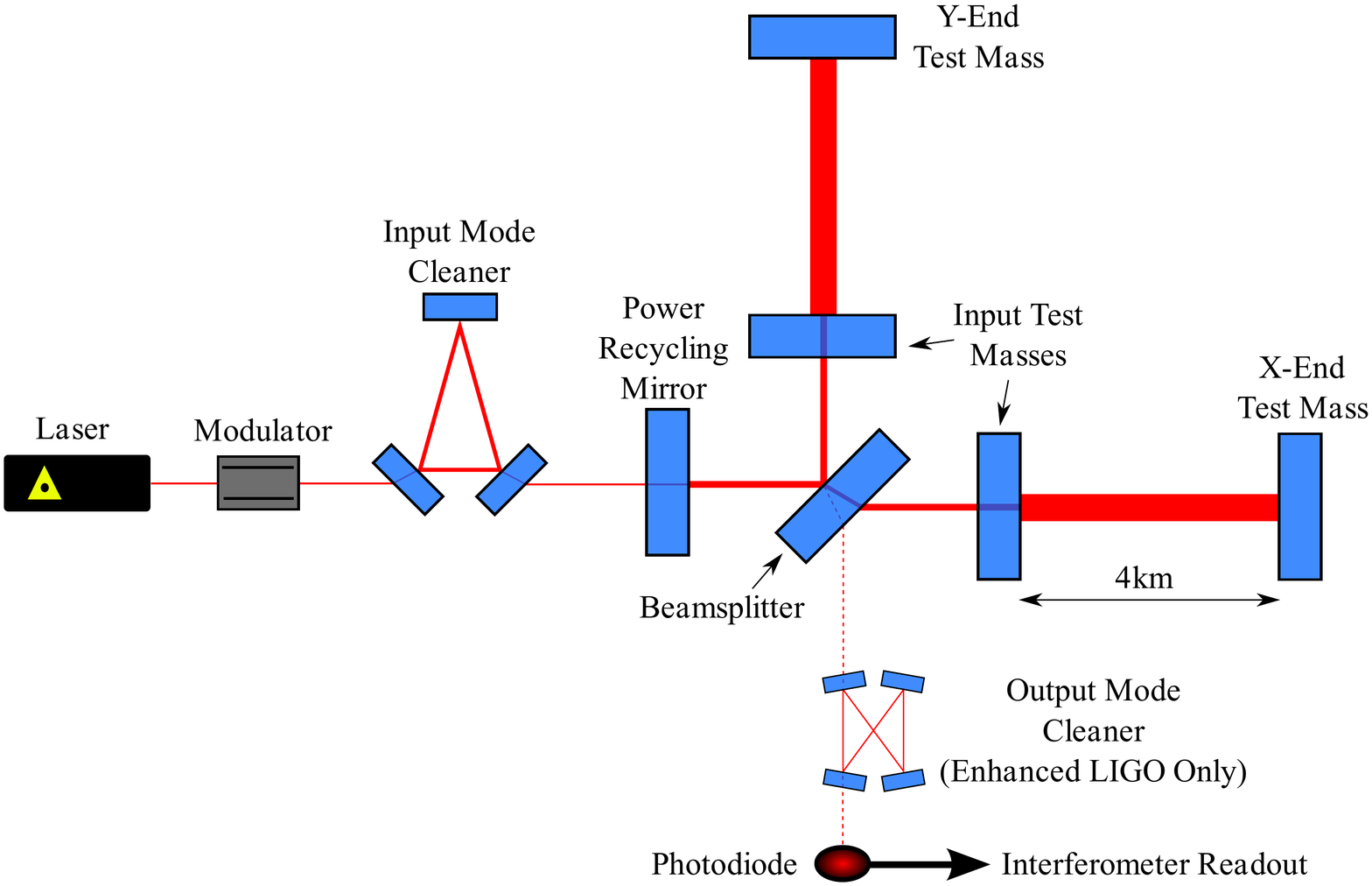}
\caption{Simplified optical layout of a 4\,km LIGO interferometer during initial LIGO (without the output mode cleaner) and Enhanced LIGO (with the output mode cleaner).}\label{fig:layout_eligo}
\end{center}\end{figure}

A thorough description of the initial LIGO instruments is given in~\cite{ligo, ligo2}. Here we give a brief overview of the instrumental configuration and sensitivity of the detectors during S5, the most recent LIGO data run. Figure~\ref{fig:layout_eligo} shows the optical configuration of an initial LIGO interferometer. The main input laser is a solid-state diode-pumped Nd:YAG system that provides roughly 10\,W of frequency and amplitude stabilized continuous wave laser light in a TEM$_{00}$ Gaussian spatial mode at 1064\,nm~\cite{ligo_laser, ligo_pstab}. The beam passes through an electro-optic modulator which adds sidebands to the carrier light at an RF offset frequency. The modulated beam is then passed through a suspended triangular cavity with a 24\,m path length and a finesse of 1350. This input mode cleaner filters higher order spatial modes from the laser beam and stabilizes its position, pointing and frequency. The beam exiting the mode cleaner is incident on the main interferometer. 

The main interferometers of LIGO can be thought of as enhanced versions of a simple Michelson interferometer, consisting of a 50/50 beam-splitter and two end test masses. If the difference of the length of the arms is held such that the light returning from the end test masses destructively interferes at the beamsplitter (the ``dark fringe'' operating point) then nominally no light exits the beamsplitter in the direction of the output photodiode.  If a gravitational wave passes and produces a differential strain on the two arms then light exits the beamsplitter and produces measurable current at the output photodiode. The optimal length of the arms is one quarter of the wavelength of a given gravitational-wave signal (so that the light spends half a period in the arms), e.g., about 2000\,km for a gravitational wave at 150\,Hz. The LIGO interferometers have only 4\,km long arms, but achieve a roughly 100 times longer effective arm length by using Fabry-Perot cavities, formed by the input test masses and end test masses, to store light in the arms. Another technique, referred to as ``power recycling'' makes use of the fact that the interferometer, when operated at the dark fringe, acts as a compound mirror for the input laser light. By installing a power recycling mirror between the input mode cleaner and beamsplitter an optical cavity is formed that effectively increases the input laser power by a factor of about 40. This is important because the sensitivity of the detector in the shot-noise-limited regime increases as the square root of the input laser power. 

The main optics of LIGO are roughly 10\,kg cylinders of high-quality fused silica, coated with quarter-wavelength stacks of dielectric materials to achieve the desired optical properties. To isolate the interferometer from seismic vibrations, each optic is suspended as a single pendulum by a loop of steel piano wire from a rigid support structure. These structures are mounted on four-layer passive vibration isolation platforms. To minimize the effects of gas on the measurements, the main optics and the 4\,km laser beam paths are enclosed in a high vacuum system.

The length of the various degrees of freedom of the interferometers, including changes in the difference of the length of the arms (the main gravitational-wave readout), are monitored using RF modulation/demodulation techniques~\cite{ligo_lsc}. The angular degrees of freedom are also monitored by a combination of RF wavefront sensing and optical levers~\cite{ligo_aa}. To keep the interferometer at a stable operating point and reduce the coupling of noise to the interferometer output, the position and orientation of each optic are controlled by actuators consisting of magnets attached to the back of the optics and coils mounted on the adjacent support structures. Many tens of additional servos are utilized to stabilize other variables, such as power and frequency of the laser. A CO$_2$ laser system is used to compensate thermal deformations of the input test masses caused by laser light absorbed by the optical coatings~\cite{tcs}.  

\begin{figure}[ht]\begin{center}
\includegraphics[width=1\textwidth, angle=0]{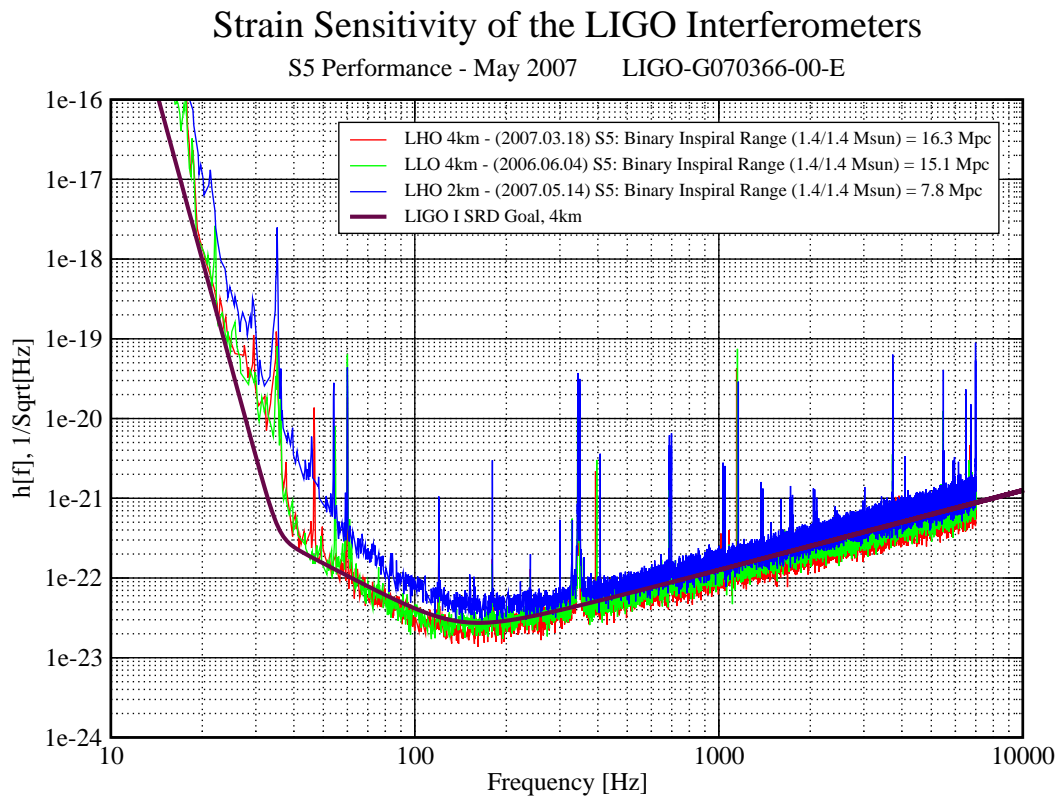}
\caption{Measured sensitivity of the initial LIGO interferometers during S5 in strain amplitude spectral density, compared to the design sensitivity goal for an initial LIGO interferometer with 4\,km arms. Also shown are the calculated distances to which each instrument could detect the inspiral of a 1.4/1.4$M_\odot$ binary system of neutron stars.}\label{fig:s5_sensi}
\end{center}\end{figure}

The sensitivity of the initial LIGO interferometers during S5 is shown in Figure~\ref{fig:s5_sensi}. The noise limit was set at low frequencies (below 40\,Hz) by seismically-driven motion of the suspended test masses and noise associated with the servo control of the angle and length degrees of freedom of the main interferometers. At high frequencies (above 200\,Hz), the limit was set by shot noise of the light detected at the interferometer outputs. At intermediate frequencies (40--200\,Hz)  a combination of sources contributed to the noise limit. Among these were length control noise, thermal noise of the main optics and their suspensions and noise from a non-linear Barkhausen effect (see, e.g., \cite{bark}) in the magnet actuators.

\section{Enhanced LIGO}\label{secn:eligo}

\begin{figure}[ht]\begin{center}
\includegraphics[width=1\textwidth, angle=0]{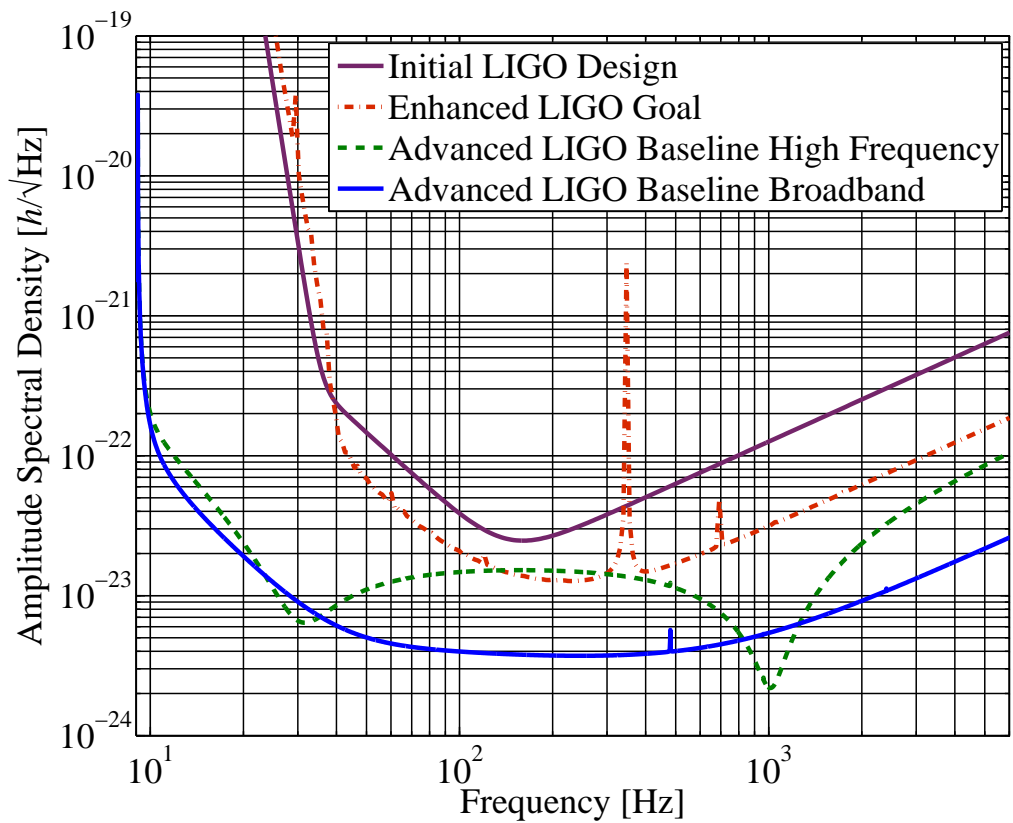}
\caption{Calculated LIGO sensitivity curves in strain amplitude spectral density. The initial LIGO curve corresponds to the ``LIGO I SRD Goal'' curve shown in Figure~\ref{fig:s5_sensi}. The Enhanced LIGO sensitivity goal is an improvement by roughly a factor of two over initial LIGO above 50\,Hz. Advanced LIGO will have a tunable sensitivity shape due to signal recycling. Two baseline curves are shown, one with ``broadband'' sensitivity and one tuned to have maximum sensitivity at higher frequency (around 1\,kHz), where the gravitational-wave radiation from many spinning neutron stars and low-mass X-ray binaries is expected to be. The initial LIGO, Enhanced LIGO and Advanced LIGO broadband curves correspond to distances to which the coalescence of a binary system of 1.4$M_\odot$ neutron stars could be detected by a single detector of roughly 15, 30, and 200\,Mpc, respectively.}\label{fig:h_i_e_a}
\end{center}\end{figure}

Enhanced LIGO is a program of moderate upgrades that is aimed at doubling the sensitivity of the two 4\,km LIGO interferometers and providing tests of several subsystems that will be key to the success of Advanced LIGO. A more detailed description of Enhanced LIGO is given in~\cite{eligo}. Installation and commissioning of new hardware for H1 and L1 began after S5 in late 2007 and is now (February 2009) nearly complete.  The first Enhanced LIGO data run, S6, is planned for mid-2009, together with Virgo. The sensitivity goal for this run is shown in Figure~\ref{fig:h_i_e_a}.  In parallel with these upgrades, the H2 detector, together with GEO600, has been operated in so-called ``astrowatch mode'' to provide observational coverage from the end of S5 to the start of S6.

To improve upon the shot-noise limit to sensitivity achieved with initial LIGO, the Enhanced LIGO interferometers use more powerful laser systems that increase the available input power from 10\,W to 35\,W. To accommodate this extra power more robust modulators and Faraday isolators and a more powerful CO$_2$ laser system to counteract the larger deformations of the input test masses are used. The gravitational-wave readout has been switched from one using RF demodulation to a so-called ``DC readout'' in which the interferometers are operated slightly ($\approx$ 10\,pm) off a dark fringe and the differential length changes of the arms result in proportional changes of the light power at the output port~\cite{dc40m}.  Other degrees of freedom are still measured and controlled as in initial LIGO, using RF readout. An output mode cleaner with a four-mirror bow-tie configuration has been installed in vacuum at the interferometer output ports to improve the DC readout signal by eliminating light in higher order spatial modes and RF sidebands that do not contribute to the strain signal, but add shot noise. This also greatly reduces the effects of phase noise and wavefront distortion of the RF sidebands on the gravitational-wave readout. The output mode cleaner is suspended as a double pendulum from a one-stage active seismic isolation system to reduce its motion. Both the output mode cleaner and its seismic isolation system are prototypes for key Advanced LIGO subsystems.

To improve the sensitivity below 200\,Hz the magnets used for test-mass actuation, which were made from NdFeB for initial LIGO, were replaced with magnets made of SmCo, a material that exhibits significantly less Barkhausen noise. Also work has been done to reduce the noise levels of electronic sensing and control signals for the main interferometer, sources of scattered light have been sought out and reduced, and the stability of the frequency of the laser has been improved. 

\section{Advanced LIGO}\label{secn:aligo}


Advanced LIGO is a major upgrade to LIGO. It will consist of three interferometers with 4\,km arms (H2 is extended), each of which will have a sensitivity roughly ten times that of an initial LIGO 4\,km detector. The interferometers will occupy the same buildings and vacuum systems as initial LIGO, but will otherwise comprise completely new hardware. Construction of Advanced LIGO subsystems has already begun and installation and commissioning will commence at the LIGO sites after S6, in 2011, with the goal of having first data collection as soon as 2014. Advanced LIGO will be sensitive enough to make routine detections of gravitational waves, and together with Advanced Virgo~\cite{avirgo, avirgo2} and GEOHF~\cite{geohf} will begin an era of gravitational-wave astronomy. Advanced LIGO is described in more detail in~\cite{aligo}.

\begin{figure}[ht]\begin{center}
\includegraphics[width=1\textwidth, angle=0]{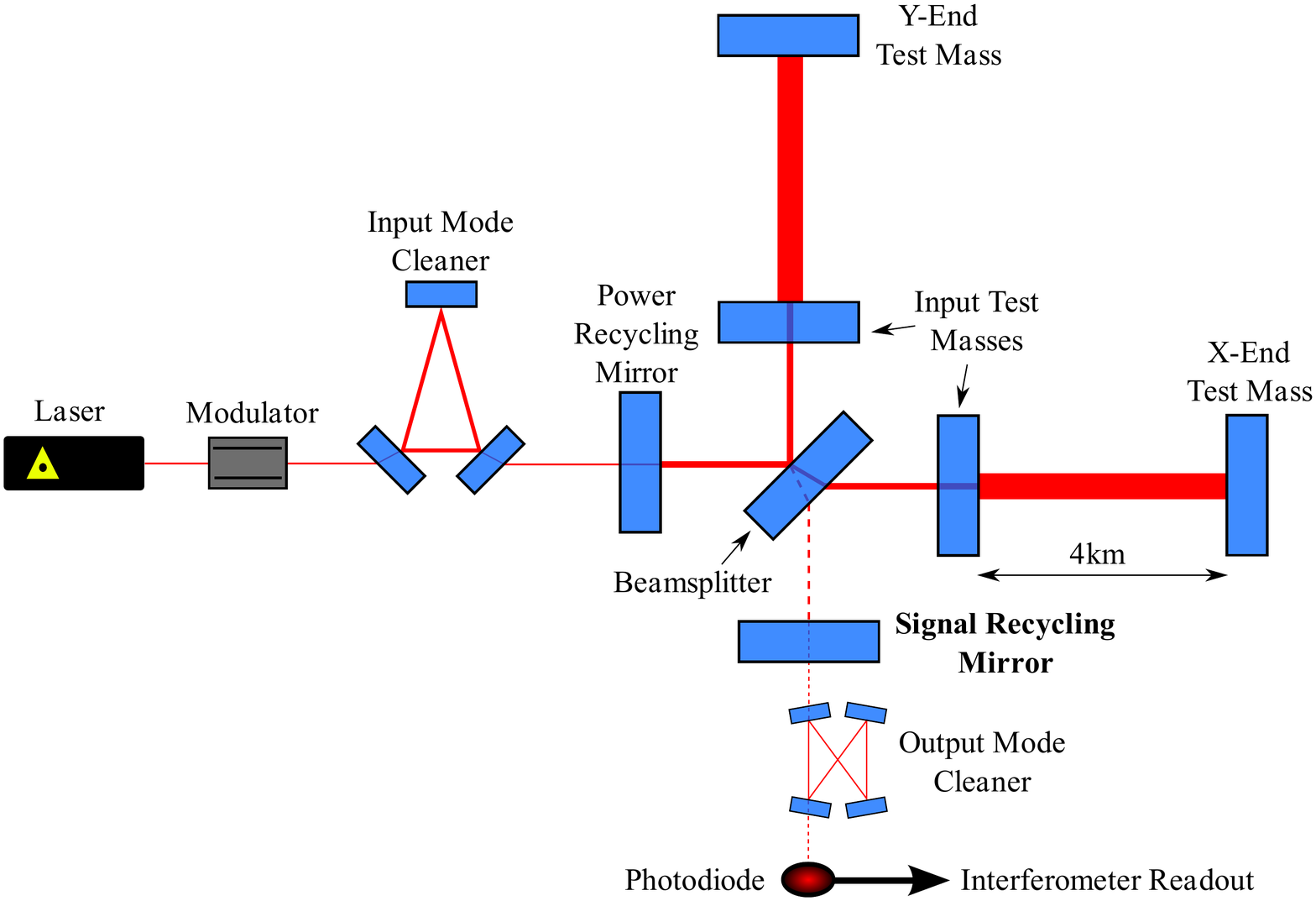}
\caption{Simplified optical layout of an Advanced LIGO interferometer. A key new addition is the signal-recycling mirror at the output port.}\label{fig:layout_adv}
\end{center}\end{figure}

The optical configuration of an Advanced LIGO interferometer is shown in Figure~\ref{fig:layout_adv}. The main interferometer is a power-recycled Michelson interferometer with Fabry-Perot arm cavities, as for initial LIGO, but has an additional mirror at the interferometer output. This ``signal-recycling mirror'' forms an optical cavity with the interferometer, allowing the gravitational-wave signal sidebands to be re-injected and stored in the interferometer or extracted depending on the cavity resonance condition.  Signal recycling~\cite{meersdr} allows for ``tuning'' of the shape of the shot-noise limited sensitivity curve of the interferometers, a technique that can be used to optimize the detector sensitivity for detecting expected astrophysical sources. Figure~\ref{fig:h_i_e_a} shows two possible sensitivity curves (only principal noise sources are included) for Advanced LIGO, one with ``broadband'' sensitivity, and another tuned to observe spinning low-mass X-ray binaries and spinning neutron stars around 1\,kHz. 

To achieve improved shot noise limited sensitivity, the input laser power will be increased to about 180\,W and the finesse of the arm cavities will be increased resulting in a circulating power in the arms of about 800\,kW compared to about 10\,kW for initial LIGO. Advanced LIGO will use the modulators and isolators tested on Enhanced LIGO. An active system will be installed to compensate any thermal deformation of the test masses due to the high laser power. 

The main optics of the Advanced LIGO interferometers will be suspended as the lowest masses of four pendulum stages~\cite{advlsus}. Behind these suspensions another four stage pendulum will be installed to provide a low noise reference for applying actuation to the test mass pendulum. The magnetic actuators used in initial LIGO will be replaced by electrostatic drives. Each test mass suspension will be mounted on a vibration isolation platform with three active stages. The effect of these systems will be to reduce the seismic cutoff frequency from 40\,Hz in initial LIGO to 10\,Hz for Advanced LIGO. 

Radiation pressure becomes an important source of noise for the levels of power in Advanced LIGO. To counteract that, more massive (roughly 40\,kg) test masses will be installed. The larger diameter of these test masses will also allow for larger beam radii, thereby reducing thermal noise from the test mass coatings and bulk materials.  To reduce thermal noise from the suspension materials, the test masses will be suspended monolithically from the penultimate masses via fused silica fibers attached to both masses with hydroxy-catalysis bonds. This system will have a higher mechanical quality factor than the steel wire slings used for initial LIGO and consequently lower off-resonance thermal noise.

\section{Summary}


Initial LIGO has finished a two year long data run with its three interferometers operating at their design sensitivity. The LIGO Scientific Collaboration is completing joint analysis of this data with the Virgo Collaboration. Following the S5 data run GEO600 and H2 have continued running in ``astrowatch mode'', to observe for gravitational waves until the LIGO and Virgo detectors are upgraded. The Enhanced LIGO upgrades to H1 and L1 are nearly complete. The next LIGO data run, S6, is planned for mid-2009 with about twice the sensitivity of S5. In parallel, construction of Advanced LIGO has begun. Installation and commissioning of Advanced LIGO at the LIGO sites will commence at the conclusion of S6 in 2011. Advanced LIGO data collection could begin as early as 2014. When fully commissioned Advanced LIGO will be sensitive enough to make multiple gravitational wave detections per year. 

\section{Acknowledgements}

The authors gratefully acknowledge the support of the United States
National Science Foundation for the construction and operation of the
LIGO Laboratory and the Science and Technology Facilities Council of the
United Kingdom, the Max-Planck-Society, and the State of
Niedersachsen/Germany for support of the construction and operation of
the GEO600 detector. The authors also gratefully acknowledge the support
of the research by these agencies and by the Australian Research Council,
the Council of Scientific and Industrial Research of India, the Istituto
Nazionale di Fisica Nucleare of Italy, the Spanish Ministerio de
Educaci\'on y Ciencia, the Conselleria d'Economia, Hisenda i Innovaci\'o of
the Govern de les Illes Balears, the Royal Society, the Scottish Funding 
Council, the Scottish Universities Physics Alliance, The National Aeronautics 
and Space Administration, the Carnegie Trust, the Leverhulme Trust, the David
and Lucile Packard Foundation, the Research Corporation, and the Alfred
P. Sloan Foundation.


\end{document}